\documentclass[twocolumn,showpacs,preprintnumbers,amsmath,prl,amssymb,superscriptaddress]{revtex4-1}

\pdfoutput=1

\usepackage[export]{adjustbox}
\usepackage{amsfonts}
\usepackage{amsmath}
\usepackage[makeroom]{cancel}
\usepackage[english]{babel}
\usepackage[T1]{fontenc}
\usepackage{times}
\usepackage{mathrsfs}
\usepackage{graphicx}% Include figure files
\usepackage{dcolumn}% Align table columns on decimal point
\usepackage{bm}% bold math
\usepackage{wasysym}
\usepackage[colorlinks,bookmarks=true,citecolor=blue,linkcolor=red,urlcolor=blue]{hyperref}
\usepackage[tight, FIGTOPCAP, hang, raggedright, nooneline]{subfigure}
\usepackage{hyperref}
\usepackage{xcolor}
\usepackage{epsfig}
\usepackage{amssymb}
\usepackage{lipsum}

\newcommand{\subfigimg}[3][,]{%
  \setbox1=\hbox{\includegraphics[#1]{#3}}% Store image in box
  \leavevmode\rlap{\usebox1}% Print image
  \rlap{\hspace*{0pt}\raisebox{\dimexpr\ht1+0\baselineskip}{#2}}% Print label
  \phantom{\usebox1}% Insert appropriate spcing
  }
  
 \definecolor{Green}{RGB}{80,182,0}

\usepackage{color}

\newcommand{\la}{\langle}
\newcommand{\ra}{\rangle}

\graphicspath{./Images/}
\usepackage{epstopdf}

\begin{document}
\title{
Absence of Ergodicity without Quenched Disorder: \\
from Quantum Disentangled Liquids to Many-Body Localization}
\author{A.~Smith}
\email{as2457@cam.ac.uk}
\affiliation{T.C.M. group, Cavendish Laboratory, J.~J.~Thomson Avenue, Cambridge, CB3 0HE, United Kingdom}
\author{J.~Knolle}
\affiliation{T.C.M. group, Cavendish Laboratory, J.~J.~Thomson Avenue, Cambridge, CB3 0HE, United Kingdom}
\author{R.~Moessner}
\affiliation{Max Planck Institute for the Physics of Complex Systems, N\"{o}thnitzer Str. 38, 01187 Dresden, Germany}
\author{D.~L.~Kovrizhin}
\affiliation{Rudolf Peierls Centre for Theoretical Physics, 1 Keble Road, Oxford, OX1 3NP, United Kingdom}
\affiliation{NRC Kurchatov institute, 1 Kurchatov sq., 123182, Moscow, Russia}
\date{\today}

\begin{abstract}
We study the time evolution after a quantum quench in a family of models whose degrees of freedom are fermions coupled to  spins, where quenched disorder appears neither in the Hamiltonian parameters nor in the initial state. Focussing on the behaviour of entanglement, both spatial and between subsystems, we show that the model supports a state exhibiting combined area/volume law entanglement, being characteristic of the quantum disentangled liquid. This behaviour appears for one set of variables, which is related via a duality mapping to another set, where this structure is absent. Upon adding density interactions between the fermions, we identify an exact mapping to an XXZ spin-chain in a random binary magnetic field, thereby establishing the existence of many-body localization  with its logarithmic entanglement growth in a fully disorder-free system. 
\end{abstract}

\maketitle

The intriguing problem of the interplay between interactions and disorder in a quantum system has been fuelling research in this field since Anderson's original work \cite{Anderson1958}. Recent progress in understanding physical phenomena associated with this interplay \cite{Gornyi:2005xy,Basko2006,Altshuler:1997fk} has firmly placed many-body localization (MBL) ideas among the central paradigms of many-body physics \cite{Altman_review2015,Nandkishore2015}. These exciting developments moved disordered interacting systems into the focus of attention, not least because MBL offers new important insights into the fundamental questions of ergodicity and its breaking, such as concepts of eigenstate thermalization hypothesis~\cite{Srednicki1994, Pino2016,Pino2017}, beyond the realm of integrable models. Because the presence/absence of ergodicity defines the way a generic system relaxes towards an equilibrium state, there are many interesting connections between the physics of MBL, and non-equilibrium quantum physics, e.g.~quantum quenches.

One of such connections, recently proposed theoretically~\cite{Grover2014,Veness2016,Garrison2017}, suggests a new non-ergodic state of matter -- the quantum disentangled liquid (QDL) -- which complements the established phenomenology of relaxation in isolated many-body quantum systems. The defining feature of these quantum liquids is that they are unable to fully thermalize because of interactions, thus making unnecessary the usual requirements for ergodicity breaking, such as integrability or quenched disorder. The idea of QDLs can be traced back to early works of Kagan and Maksimov on interaction-induced localization, discussed in the context of solid Helium~\cite{Kagan}. One QDL scenario is that of heavy particles which thermalize, while light particles evade thermalization by localizing on the heavy particles. More recent studies of heavy-light particle models suggest that this physical picture of sub-diffusive dynamics, while present, is only transient, and gives way to ergodic behaviour at long times. Hence, these systems have been dubbed quasi-MBL~\cite{Yao2014,Papic2015,Schiulaz2015}. Similar phenomenology has been observed in the corresponding quantum dynamics of classical glassy models~\cite{vanHorssen2015,Hickey2016}.  Intriguingly, some evidence for QDL-like behaviour, showing different timescales for equilibration of two subsystems, has been observed in cold-atom experiments~\cite{Bloch_private}.

In a recent paper~\cite{Smith2017} we proposed a disorder-free spin-fermion model, which exhibits complete localization of the fermion subsystem. Its remarkable feature is that disorder, a prerequisite for localization, only emerges dynamically. This is highlighted via an exact duality mapping between spin/fermion degrees of freedom. This non-linear transformation reveals the presence of an extensive number of conserved quantities playing the role of the disorder potential. In the dual representation the model becomes that of free-fermions, and there is an important question as to what extent the physics that we found is robust to adding perturbations to our model. Here we propose and study such an interacting extension, showing that it can be mapped exactly onto a random field XXZ spin-chain -- the drosophila of MBL~\cite{Znidaric2008,Bardason2012,Enss2016,Schreiber2015,Andraschko2014,Tang2015}.

A standard diagnostic for MBL and QDL behaviour is the bipartite entanglement entropy. Many-body localization can be distinguished from its non-interacting counterpart -- Anderson localization~\cite{Anderson1958} -- via the post-quench logarithmic growth of entanglement compared with the area-law saturation of entanglement correspondingly~\cite{Znidaric2008,Bardason2012}. QDLs on the other hand can be identified using projective measures of entanglement entropy of separate species~\cite{Grover2014}. One of these obeys an area-law scaling, while the other together with the full system show the volume-law. The original proposal of~\cite{Grover2014} provides explicit examples of many-body wave functions showing QDL phenomenology. However, the search for a microscopic Hamiltonian supporting quantum disentangled liquid has so far proved to be inconclusive. In this Letter we demonstrate two central results obtained within our model; the many-body localization without quenched disorder, and a microscopic Hamiltonian showing QDL behaviour. Here we focus on the results for the time evolution of entanglement entropy after a quantum quench, which are obtained using a combination of duality mappings, exact diagonalization, and matrix-product state (MPS) based time evolution.

Our work comes at a time when exceptional progress has been made in experimental realization of controlled isolated quantum systems~\cite{Schreiber2015,Choi2016,Zhang2017} and in simulating lattice gauge theories coupled to fermionic matter~\cite{Martinez2016} -- of which our system is an example. This is driven in part by MBL and general questions about thermalization, or lack-thereof, in such systems. The Hamiltonian we present is simple enough that it should be implementable in similar set-ups, and being able to tune the localization length should minimize the effect of system size limitations. We have a system that violates the eigenstate thermalization hypothesis in the two ways that we present in this paper, and with a novel disorder-free mechanism.

%%%%%
\emph{Model and its mapping to XXZ chain in a random field.---}
In our previous work~\cite{Smith2017} we introduced a model of spinless fermions, $\hat{f}_j$, hopping between sites of a 1D lattice, that are coupled to spins-1/2, $\hat{\sigma}_{j,j+1}$, living on the bonds. Here we extend this model by adding nearest-neighbour interactions between the fermions
\begin{equation} \label{eq: H}
\begin{aligned}
\hat{H_f} = -J\sum_{\la i j\ra} \hat{\sigma}^z_{i,j} \hat{f}^\dagger_i \hat{f}_j - h\sum_j \hat{\sigma}^x_{j-1,j} \hat{\sigma}^x_{j,j+1}\\
 + \Delta \sum_j (2\hat{n}_j -1)(2\hat{n}_{j+1} - 1),
\end{aligned}
\end{equation}
where $\hat{n}_j = \hat{f}^\dagger_j \hat{f}_j$ is the fermion density operator. Without loss of generality we assume that all parameters of the Hamiltonian are non-negative. The model possesses an extensive number of conserved quantities (charges), identified by a duality mapping which we outline here for completeness, see details in~\cite{Smith2017}. We define $\tau$-spins on the sites of the lattice through the duality transformation~\cite{Kramers1941,Fradkin1978}
\begin{equation}
\hat{\tau}^z_j = \hat{\sigma}^x_{j-1,j} \hat{\sigma}^x_{j,j+1}, \qquad \hat{\tau}^x_{j} \hat{\tau}^x_{j+1} = \hat{\sigma}^z_{j,j+1}.
\end{equation}
The charges $\hat{q}_j \equiv \hat{\tau}^z_j (-1)^{\hat{n}_j}$, commute with the Hamiltonian also in the presence of fermion interactions $\Delta \neq 0$. Finally, in terms of new fermion operators $\hat{c}_j = \hat{\tau}^x_j \hat{f}_j$ the Hamiltonian can be recast in the following form
\begin{equation} \label{eq: H XXZ}
\begin{aligned}
\hat{H_q} = -J\sum_{\la i j\ra} \hat{c}^\dagger_i \hat{c}_j + h\sum_j \hat{q}_j (2\hat{n}_j - 1)\\
+ \Delta \sum_j (2\hat{n}_j -1)(2\hat{n}_{j+1} - 1),
\end{aligned}
\end{equation}
where $\hat{n}_j = \hat{c}^\dagger_j \hat{c}_j = \hat{f}^\dagger_j\hat{f}_j$, and $\hat{q}_j$ have eigenvalues $\pm 1$. The Hamiltonian~\eqref{eq: H XXZ} is equivalent to an XXZ chain in a magnetic field via Jordan-Wigner transformation \cite{Essler2016}, where the value of the magnetic field on each lattice site is given by $\pm 2h$, and the signs are fixed for any given configuration of $q_j$'s. In the following we investigate the emergence of QDL and MBL behavior using the time evolution of the entanglement entropy after a global quantum quench with the Hamiltonian ($\ref{eq: H}$), and initial states being tensor products of spin and fermion degrees of freedom. 

For simplicity we assume that at $t=0$ the $\sigma$-spins are polarized along the z-axis, and the $f$-fermions are described by the Slater determinant corresponding to a charge density wave. Thus, initial states $|0\ra = |\uparrow \uparrow \cdots\ra_\sigma \otimes |\psi\ra_f$ transform into an equal-weight superposition of charge configurations $|0\ra = \frac{1}{\sqrt{2^N}} \sum_{\{q_j\} = \pm 1} |q_1 q_2 \cdots q_N \ra \otimes | \psi \ra_c$, with $|\psi\ra_f$ equivalent to $|\psi\ra_c$~\cite{Smith2017}.
Note that the choice of a spin-polarized initial state is dictated purely by its simplicity, while the physics remains the same for any typical spin state. Exceptions are a zero-measure subset of special states, e.g., there is a simple product state of spins each (anti-)aligned with the x-axis and fermions in a tensor product of local occupations (such as a CDW) which maps to fermions in a single uniform charge sector.

In this setup the problem maps to a paradigmatic MBL system -- the XXZ spin-chain in a random magnetic field~\cite{Znidaric2008}. In our case the field has a binary nature, in other words it takes only two values $\pm 2h$, as in Refs.~\cite{Andraschko2014,Tang2015} where MBL behaviour is also observed. Note, that here disorder is determined by the conserved charges $\hat{q}_j$ which are themselves related to the physical degrees of freedom of Eq.~\eqref{eq: H}. Our choice of the initial state results in averaging over all charge configurations, thereby generating emergent random binary magnetic fields. 

%%%%%
\emph{Quantum Disentangled Liquid.---}A fresh perspective using entanglement measures~\cite{Grover2014} was recently proposed in the context of localization in a disorder-free system given by a mixture of heavy and light particles \cite{Kagan}. These developments brought forward the notion of a quantum disentangled liquid -- a state of matter which is defined by different behaviour of the entanglement entropy of its subspecies. However, to our knowledge, no microscopic Hamiltonian conclusively exhibiting this behaviour has been identified so far. Here we show that the model we suggested in \cite{Smith2017}, even in the non-interacting case of $\Delta=0$, does realize the phenomenology of QDLs.

\begin{figure}[th!]
	\centering
	\subfigimg[width=.45\textwidth]{\hspace*{0pt} \textbf{(a)}}{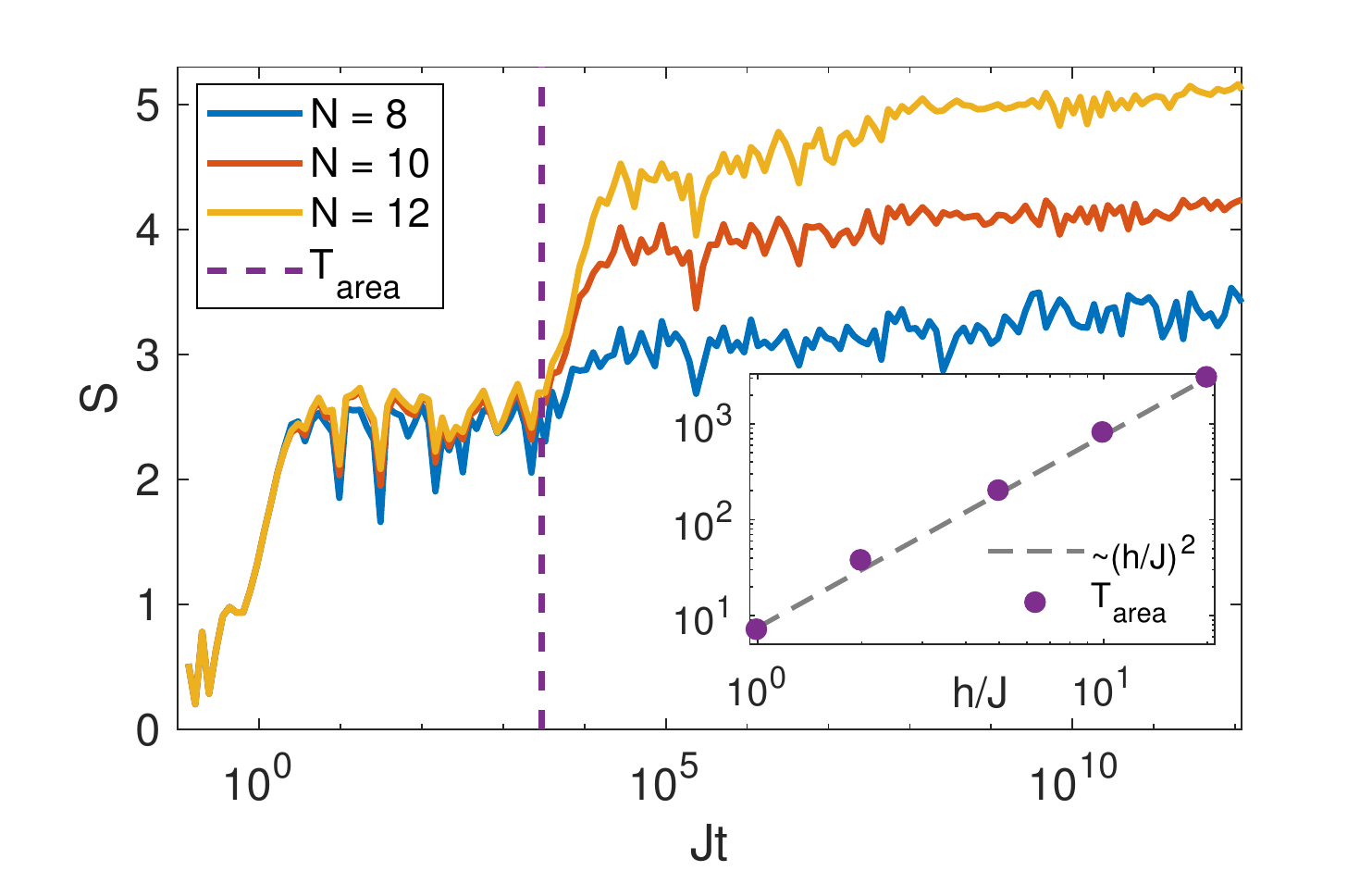}
	\!\!\!\!\subfigimg[width=.45\textwidth]{\hspace*{0pt} \textbf{(b)}}{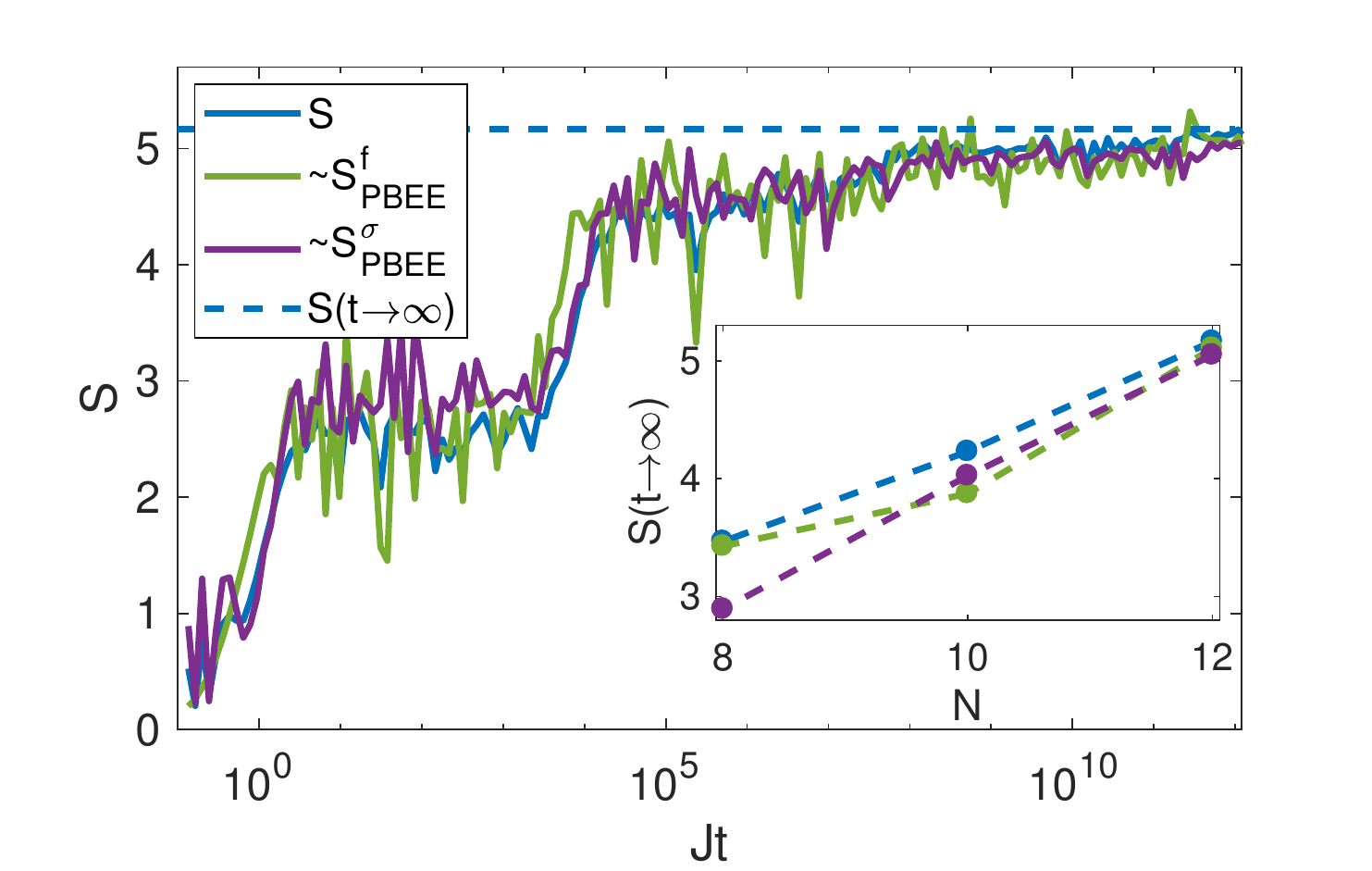}
\caption{Time evolution of entanglement entropy after a quench from a charge density wave state. The results are obtained using exact diagonalization for $h/J =20$, $\Delta=0$. (a) The von Neumann bipartite entanglement entropy $S(t)$ of a half-system for $N=8,10,12$. (inset) The time $T_\text{area}$ for which the area-law plateau  persists (dashed line of main plot) as a function of $h/J$ compared with $(h/J)^2$. (b) Comparison between PBEEs $S^f_\text{PBEE}(t)$ and $S^\sigma_\text{PBEE}(t)$ and the entanglement of the full system $S(t)$ for $N=12$. (inset) The long time-limit $S(t\rightarrow \infty)$ (computed at $Jt\sim10^{12}$) as a function of system size. PBEE results are scaled by factors of $4.3$ and $1.9$, respectively.}\label{fig: EE}
\end{figure}

The quantum disentangled liquid was defined in \cite{Grover2014} via Projective Bipartite Entanglement Entropy (PBEE). Here we briefly review the definition of PBEE for the case of a system with two components in a pure state $|\psi \ra$. Let $\alpha$ and $\beta$ label the components, and $\hat{P}^\gamma_\phi$ be a projector onto the state $|\phi\ra$ of the species $\gamma \in \{\alpha, \beta\}$. This projector is related to a measurement of the single component. We also spatially partition our system into two subsystems, $A$ and $B$. The algorithm for calculating PBEE for the component $\alpha$ is as follows: 
(i) project the state $|\psi\ra$ onto the state $|\phi\ra$ of species $\beta$, i.e. $|\psi\ra_\phi = P^\beta_\phi | \psi \ra$; 
(ii) define the reduced density matrix $\rho^\phi_A = \mathrm{Tr}_B |\psi\ra_\phi  \la \psi |_\phi$;
(iii) compute the von Neumann entanglement entropy $S^\phi_A = -\text{Tr}_A[\rho^\phi_A \log \rho^\phi_A]$;
(iv) the PBEE for the species $\alpha$ is then defined as
\begin{equation}
S^\alpha_\text{PBEE} = \sum_\phi ||\psi\ra_\phi|^2 S^\phi_A,
\end{equation}
where the sum for the entropies $S^\phi_A$ is weighted with the probabilities of states $|\psi\ra_\phi$. Crucially a QDL has volume-law scaling of the total bipartite von Neumann entropy $S$ and $S^\alpha_\text{PBEE}$ for one species, but the area-law for the other species $S^\beta_\text{PBEE}$. 

\begin{figure}[b]
	\centering
	\subfigimg[width=.45\textwidth]{\hspace*{0pt} \textbf{(a)}}{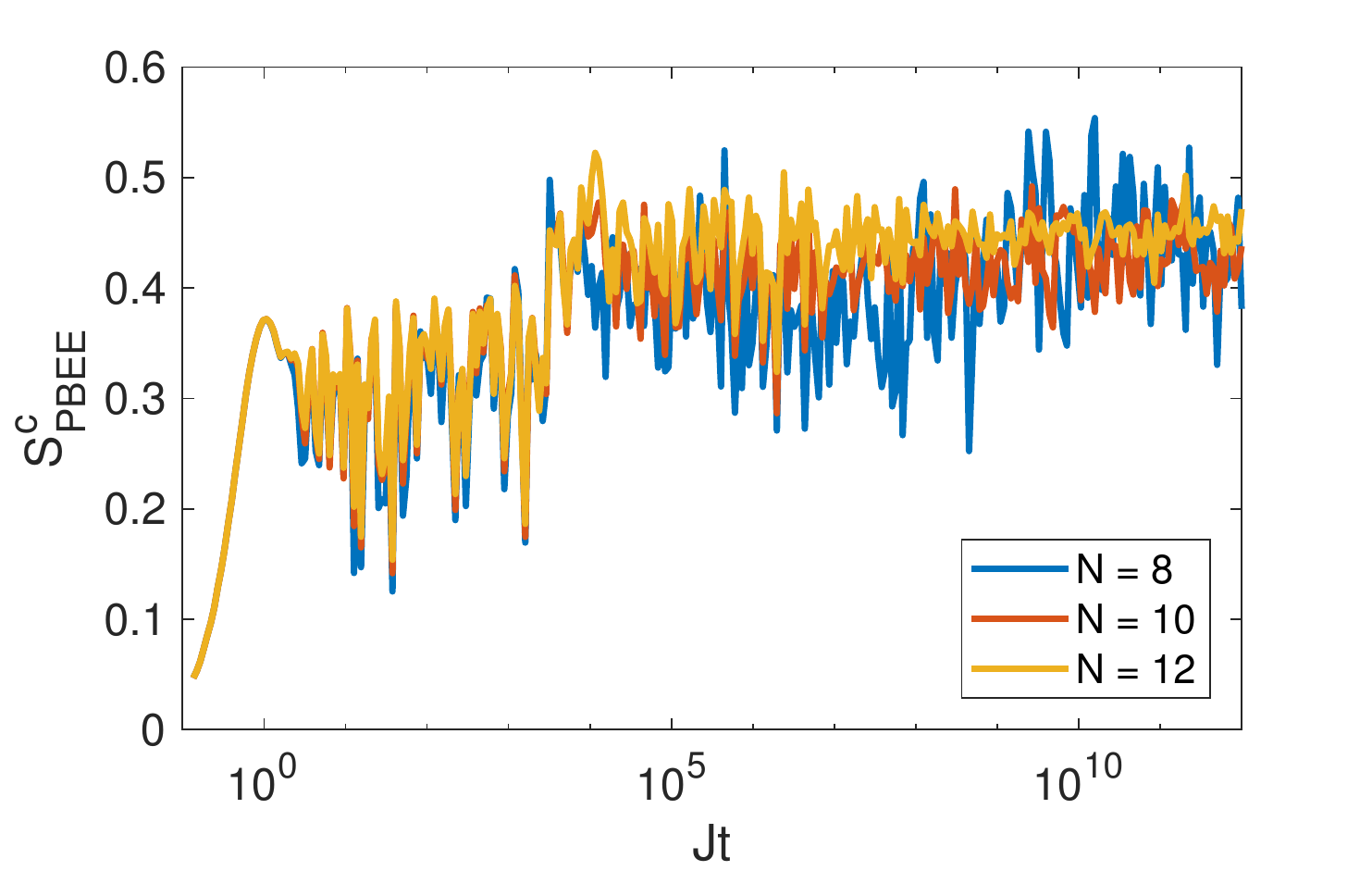}
	\subfigimg[width=.45\textwidth]{\hspace*{0pt} \textbf{(b)}}{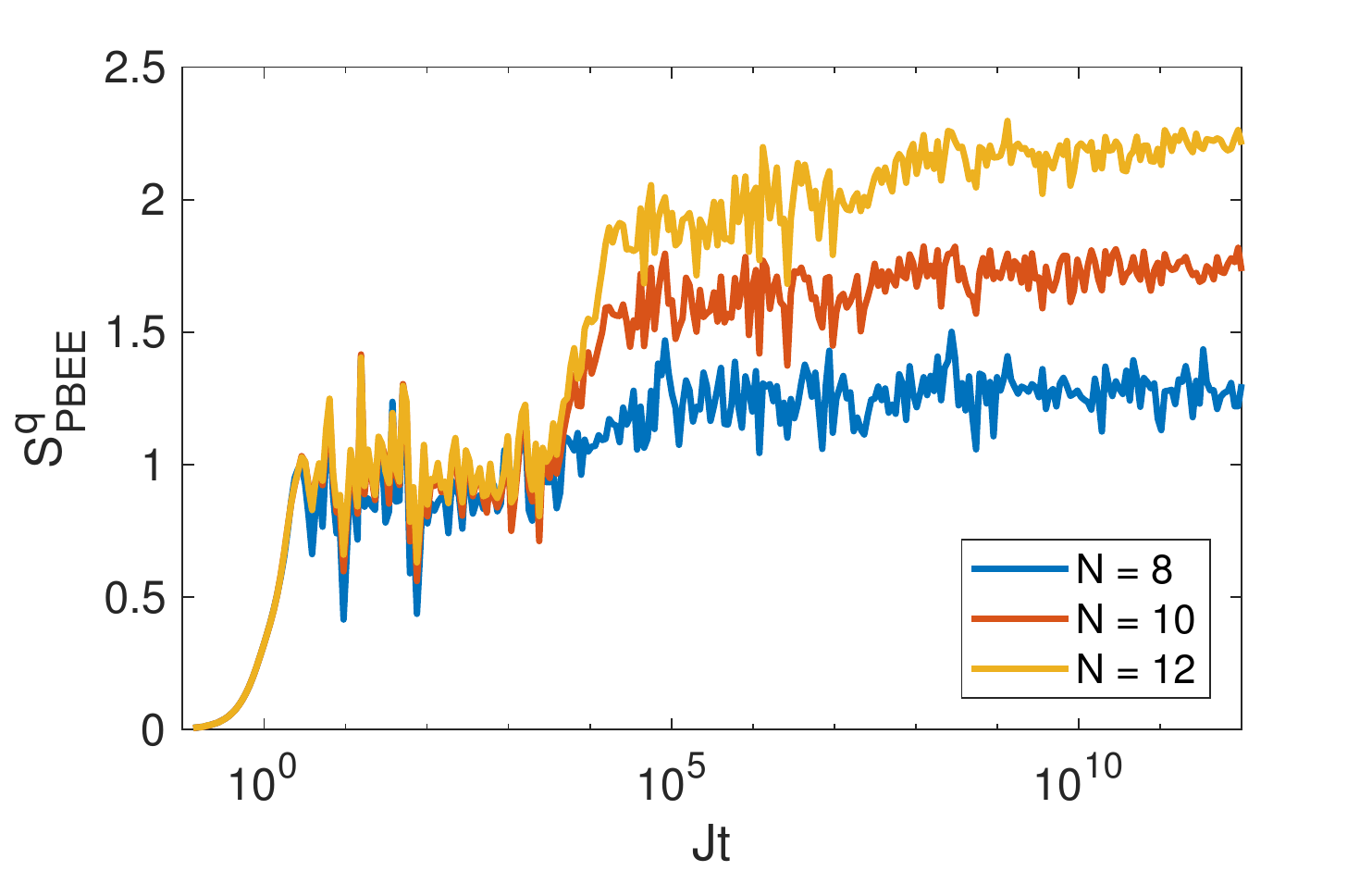}
\caption{The time evolution of the PBEE starting from a charge density wave state for $h/J=20$, $\Delta=0$, and $N=8,10,12$, obtained using exact diagonalization. (a) $S^c_\text{PBEE}(t)$ for $c$-fermions. (b) $S^q_\text{PBEE}(t)$ for the conserved charges $q_i$, see text.}\label{fig: PBEE cq}
\end{figure}

In Fig.~\ref{fig: EE}(a) we show bipartite entanglement entropy for the full system for $\Delta=0$, $h/J = 20$ after a quench from a charge density wave fermion state. The entropy exhibits initial linear growth followed by an area-law plateau which eventually gives way to the volume-law scaling (note the dependence on the system size). The extent of the plateau scales as $(h/J)^2$  for $h/J \gg 1$, as shown in the inset; it is absent for $h/J < 1$. This behaviour can be attributed to a separation of timescales, which is particularly crisp in our case of binary disorder, where for $h/J \gg 1$, a pair of adjacent sites with opposite
values of $q_j$ correspond to a high energy barrier. Traversing such a barrier is a process parametrically suppressed in $h/J$, while motion
between such barriers takes place on shorter timescales. The latter can only produce  area-law scaling of the entanglement entropy, while the former
can act on longer timescales, resulting in equilibration of the spins and a concomitant 
volume-law scaling for the  entanglement entropy. Note that the same two localization regimes also appear in  the disorder-averaged entanglement entropy of a simple tight-binding model with binary disorder. It is directly related to PBEE projected onto the charge sectors in our model, $S^c_\text{PBEE}$ shown in Fig.~\ref{fig: PBEE cq}(a), because our choice of spin polarized initial state leads to an equal weight superposition of all disordered charge configurations.

The PBEEs for the original degrees of freedom, the $f$-fermions and $\sigma$-spins, are shown in Fig.~\ref{fig: EE}(b). The data is scaled to highlight the fact that both PBEEs have the same qualitative behaviour, and match the entanglement entropy of the composite system. In terms of the $f$ and $\sigma$ degrees of freedom, the long time limit {\it does not} suggest the QDL behaviour since all three measures develop volume-law scaling (see inset). However, in terms of new degrees of freedom, after the mapping to $c$-fermions and conserved charges, we {\it do} find the phenomenology of the QDL. The corresponding PBEEs obey area and volume law scaling, respectively, as shown in Fig.~\ref{fig: PBEE cq}. Importantly, we find area-law scaling of the PBEE for a macroscopic fraction of the degrees of freedom. Furthermore, since the localization behaviour persists for all system sizes~\cite{Smith2017}, and there is a direct relation between the area-law scaling of $S^c_\text{PBEE}$ and the localization of fermions, this allows us to infer that this behaviour holds in the thermodynamic limit. 

These contrasting results highlight the subtlety of defining a QDL, most crucially an appropriate choice of the measurement basis. While the dynamics of the $f$ and $c$ fermions is closely related, e.g.,~all density correlators are the same, they are connected via non-linear and non-local transformation with a string of spin operators.

\begin{figure}[t]
	\centering
	\subfigimg[width=.42\textwidth]{\hspace*{0pt} \textbf{(a)}}{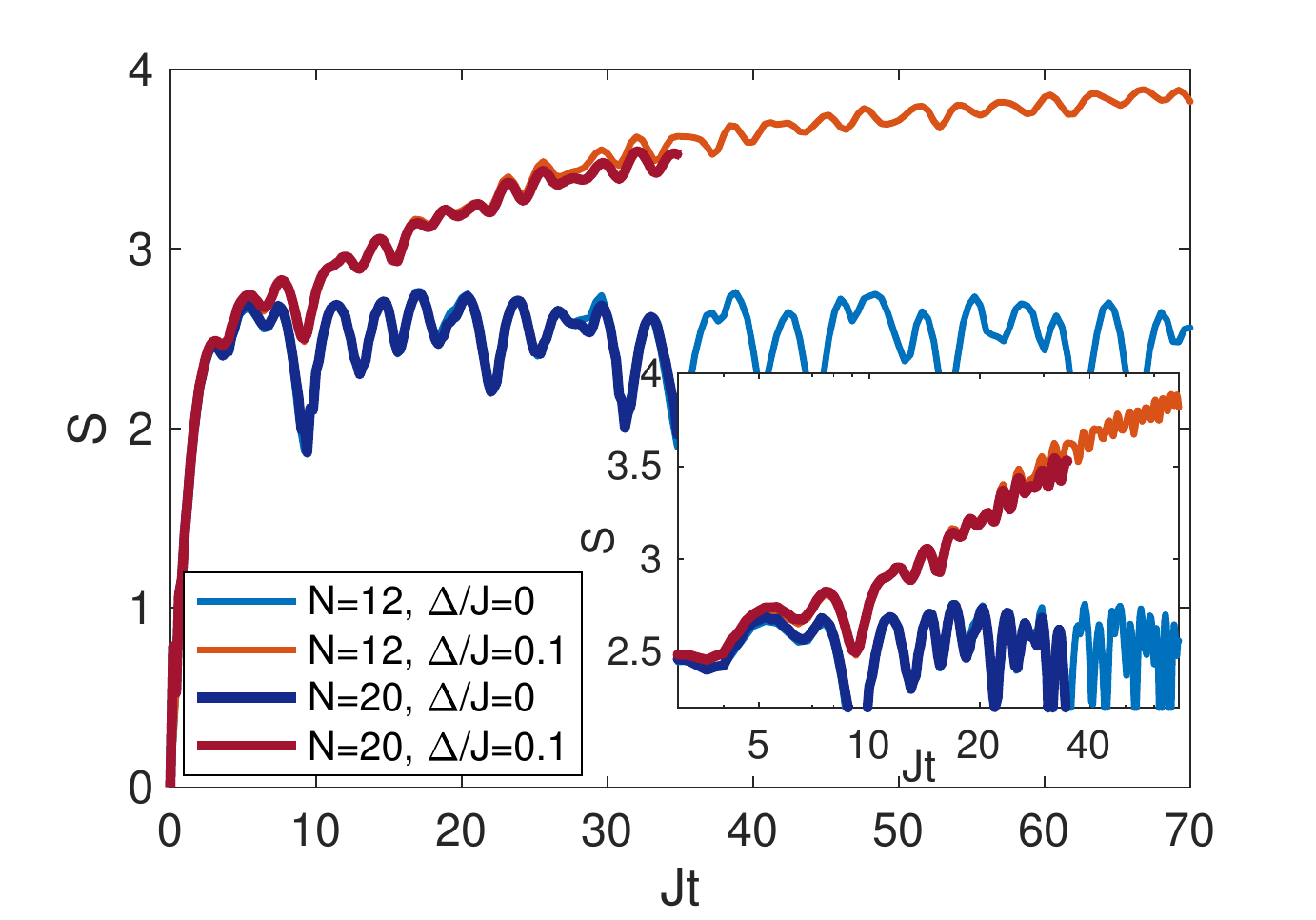}
	\!\!\!\!\subfigimg[width=.42\textwidth]{\hspace*{0pt} \textbf{(b)}}{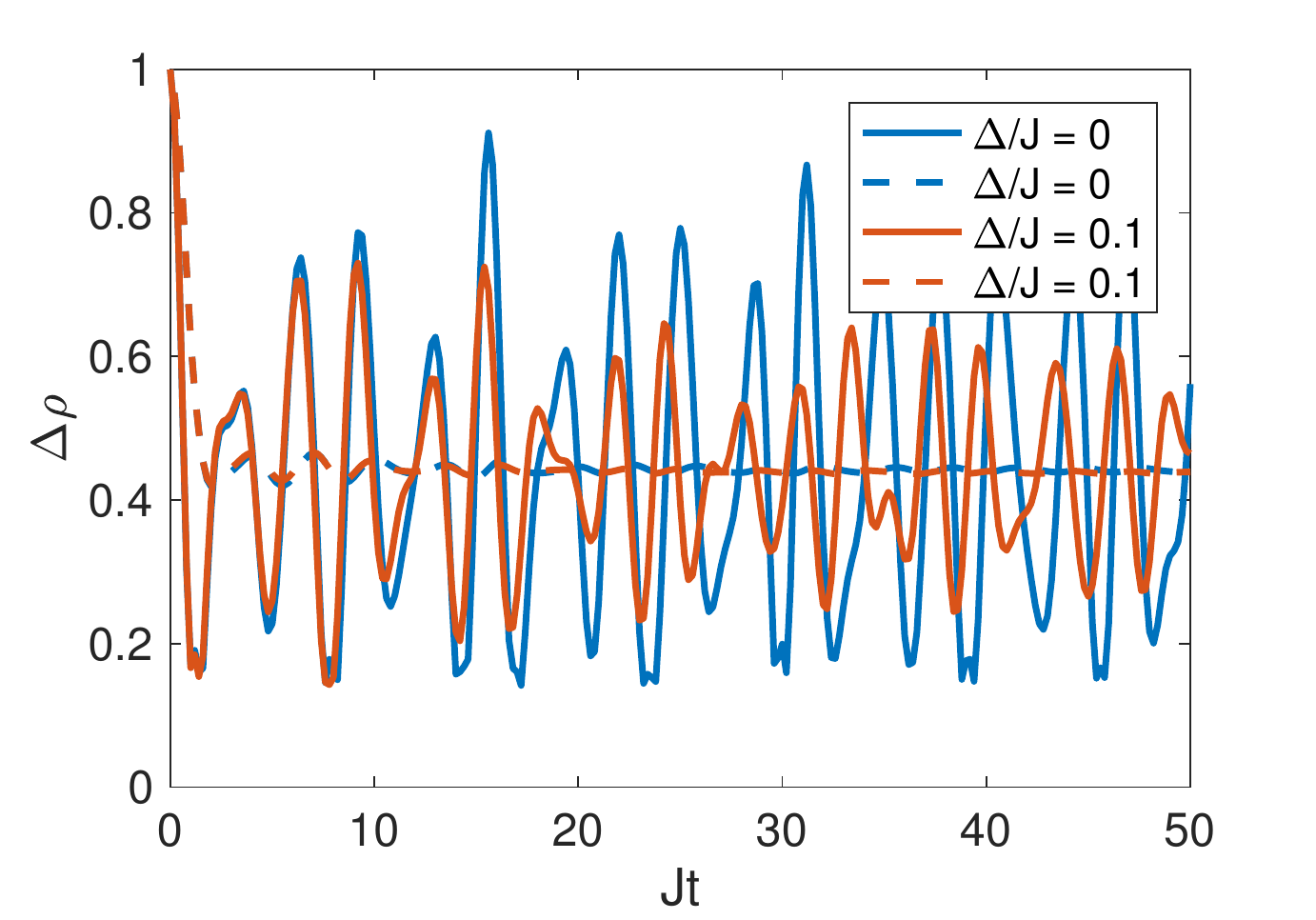}
\caption{Quantum quench from an initial charge density wave state, $h/J=20$. (a) Von Neumann entanglement entropy computed by ED for N=12 (thin, light) and an MPS algorithm for $N=20$ (thick, dark) with $\Delta/J = 0, 0.1$. The spatial bipartition is taken along the central bond. (inset) The same data on a semi-log plot. (b) Density imbalance $\Delta\rho(t)\propto \sum_j |\la 0 | \hat{n}_j(t) - \hat{n}_{j+1}(t) |0\ra|$ and the time-averaged value $\frac{1}{t}\int^t_0\text{d}\tau\,\Delta\rho(\tau)$ (dashed lines) after the same quench.}
\label{fig: MBL}
\end{figure}
	
%%%%%
\emph{Disorder-free MBL.---}We now turn to our second main result related to the interacting fermion case $\Delta \neq 0$. 
Here, the system~\eqref{eq: H XXZ} can be mapped to an XXZ model with a random magnetic field of binary nature $q_j h\to \pm h$ via a standard Jordan-Wigner transformation, $S^+_j=\hat{c}_j^{\dagger}(-1)^{\sum_{l<j}\hat{n}_l}$ and $S^z_j=\hat{n}_j-\frac{1}{2}$, yielding
\begin{multline}
\hat{H}_{XXZ} = -J\sum_{j} (\hat{S}^+_j \hat{S}^-_{j+1}+\hat{S}^-_j \hat{S}^+_{j+1})\\ + 4\Delta \sum_{j}\hat{S}^z_j\hat{S}^z_{j+1} + 2h\sum_j  q_j \hat{S}^z_j.
\end{multline}
Usually studied with continuously sampled disorder, but also considered with binary disorder in Ref.~\cite{Andraschko2014,Tang2015}, the random field XXZ model serves as an important example of a model showing many-body localized behaviour~\cite{Znidaric2008}. Here, we find that MBL phenomenology extends to our model, even without quenched disorder. MBL is often distinguished from Anderson localization by the logarithmic growth of entanglement entropy after a quench whilst preserving area-law scaling~\cite{Bardason2012,Serbyn2013} with the system size. We use this diagnostic for the initial charge density wave fermion state. The time evolution under the Hamiltonian (\ref{eq: H}) is computed using exact diagonalization for $N=12$ and by an MPS algorithm for $N=20$ (with the help of the iTensor library~\cite{iTensor}), where we use second-order Trotter decomposition with error of compression at each step less than $3\times 10^{-7}$ up to a maximum bond dimension $\chi = 700$.

In Fig.~\ref{fig: MBL}(a) we present the results for the time evolution of entanglement entropy after a quench from a charge density wave initial state. In the case of $\Delta=0$ we observe an area-law plateau at long times, as identified in Fig.~\ref{fig: EE}(a). Upon increasing $\Delta/J$ to $0.1$ we find a change of behaviour with entanglement entropy growing without saturation, which also obeys area-law scaling with respect to different size partitions (not shown) and is evident from comparing results for $N=12$ and $N=20$. The same data shown in a semi-log plot (see inset) confirms that this is consistent with the logarithmic growth of entanglement. The averaged density imbalance between neighbouring sites, $\Delta\rho(t) \propto \sum_j |\la 0 | \hat{n}_j(t) - \hat{n}_{j+1}(t) |0\ra|$, along with the time-averaged value $\frac{1}{t}\int^t_0 \text{d} \tau\, \Delta\rho(\tau)$, is shown in Fig.~\ref{fig: MBL}(b). For both $\Delta/J = 0, 0.1$, the density imbalance oscillates about the long-time value $\Delta\rho(\infty) > 0$, directly establishing non-ergodicity on these very large timescales~\cite{Smith2017}. The electron interactions ($\Delta\ne0$) lead to additional damping of the oscillations around this value. For a more detailed investigation of the XXZ spin chain with binary disorder we point the reader to Refs.~\cite{Andraschko2014,Tang2015}

%%%%%
\emph{Discussion.---}We presented an extension of our model of disorder-free localization discussed previously in Ref.~\cite{Smith2017}. The model shows rich phenomenology, from quantum disentangled liquids to many-body localization. Our results explicitly demonstrate that the usual assumption that the MBL phase requires quenched disorder is false.

Using the time evolution of entanglement after a quantum quench we demonstrated that our model also shows quantum disentangled liquid behaviour in this observable, and we highlight the dependence of the definition of QDL on the choice of the measurement basis. In spite of a close relationship between the $f$ and $c$ fermions that we consider, the projective entanglement involved in the definition of QDLs reveal a stark difference between the two measurements. We have also identified a family of models that realise QDLs, namely extensions to our model, which commute with the conserved charges, as well as the models with non-interacting auxillary spins of Ref.~\cite{Paredes2005}. 

The model can be extended by adding to the Hamiltonian a number of other terms commuting with conserved charges. A particularly interesting example is a simple longitudinal field $\sim\sum_i \sigma_i^x$. This term confines excitations of the spin sector and corresponds to a non-local interaction after the mapping to $c$-fermions. Another possible extension is to give dynamics to the conserved charges which could help to establish how robust the physics of disorder-free MBL is to perturbations.

The model we discussed in this Letter marks an intersection of many-body localization and quantum disentangled liquids. Recent experimental progress on controlled isolated quantum systems, and in particular the simulation of lattice gauge theories, makes it accessible with current capabilities~\cite{Schreiber2015,Choi2016,Zhang2017,Martinez2016}. It provides a new setting for studying old and general open questions about the relaxation of isolated many-body quantum systems.

\section*{Acknowledgements} We are grateful to Fabian Essler for enlightening and encouraging suggestions, and for bringing to our attention the importance of the XXZ mapping. We would like to thank Thomas Veness for his suggestions on the manuscript. A.~S.~acknowledges EPSRC for studentship funding under Grant No.~EP/M508007/1. J.~K. is supported by the Marie Curie Programme under EC Grant agreements No.703697. The work of D.~L.~K.~was supported by EPSRC Grant No.~EP/M007928/1.  R.~M.\ was in part supported by DFG under grant SFB 1143.

%\bibliography{library}


\end{document}